# Optimized Password Recovery for Encrypted RAR on GPUs

Xiaojing An, Haojun Zhao, Lulu Ding, Zhongrui Fan, Hanyue Wang
Institute of Computing Technology
Chinese Academy of Sciences
Beijing, China
{thefighteran, zhaohaojun, dinglulu14, fanzhongruis}@163.com, wanghanyue14@mails.ucas.ac.cn

*Abstract*—RAR uses classic symmetric encryption algorithm SHA-1 hashing and AES algorithm for encryption, and the only method of password recovery is brute force, which is very time-consuming. In this paper, we present an approach using GPUs to speed up the password recovery process. However, because the major calculation and time-consuming part, SHA-1 hashing, is hard to be parallelized, so this paper adopts coarse granularity parallel. That is, one GPU thread is responsible for the validation of one password. We mainly use three optimization methods to optimize this parallel version: asynchronous parallel between CPU and GPU, redundant calculations and conditional statements reduction, and the usage of registers optimization. Experiment result shows that the final version reaches 43~57 times speedup on an AMD FirePro W8000 GPU, compared to a well-optimized serial version on Intel Core i5 CPU.

## I. INTRODUCTION

RAR is one of the most popular compressed file formats. The encryption method of RAR is based on the powerful AES-128 standard [1]. The encryption process includes SHA-1 hashing and AES algorithm with data dependency, and both of them are currently unable to be cracked. The dictionary file, brute force and brute force with a mask are the only ways to recover the password of an encrypted RAR file. These three methods are essentially based on the same idea. They are all implemented by traversing all potential passwords, using them to decrypt the file header block, and verify the correctness by the CRC cksum. The only difference is the way to capture the test passwords. Since the calculations of all potential passwords are independent with each other, RAR password recovery algorithm has good parallelism, and is suitable to work on GPUs. However, the calculation of each password is massive, and difficult to be separated to work in parallel, so the RAR password recovery algorithm is coarse granularity parallel.

This paper proposes a series of optimized measures for the algorithm's GPU implementation:

- Asynchronous parallel between CPU and GPU;
- Reducing redundant calculations and conditional statements by pre-computing and uniformly storing the data to be assigned, to trade space for time;
- Optimizing the usage of register by moving the frequently accessed data from register to local memory, when the registers are abused;
- Others, like reusing registers, optimization of calculations, finding the most optimal number of workgroups and so on.

After the step-wise optimizations, our implementation of RAR password recovery algorithm achieves a performance speed-up between 43 and 57 times with the password length between 4 and 25, compared with AccentRAR, a mature commercial software, running on CPU.

The main contribution of this paper is to provide a complete and high-performance implementation of RAR password recovery algorithm in OpenCL. The proposed solution features the following novelties: We assign different kinds of works in RAR password recovery to different kinds of architectures, and bring the GPU and CPU into effective teamwork by the asynchronous parallel; Per-computing the same assignment calculation of all threads and uniformly storing the result to reduce and simplify the calculation, which means trading space for time; Because the program is coarse granularity parallel, which leads to the abuse of registers, we use local memory to replace some usage of the registers to avoid the frequently accessed data being moved to global memory.

This paper is structured as follows. Section II reviews the related work. Section III describes the RAR password recovery algorithm in details and analyzes its parallelism. Section IV analyzes the performance bottlenecks of the program and introduces the optimizations. Section V tests and evaluates the approaches. Section VI concludes with future work.

## II. RELATED WORK

Recent years, the GPUs have gained popularity for general purpose applications, and cryptographic technique is one of them. Powerful computational capability of GPU makes brute force algorithm available. So a lot of work has been done to accelerate brute force password recovery of RAR, PDF, DOC documents, and WIFI password by GPU. There are many mature commercial softwares of the brute force password recovery of RAR, such as AccentRAR and cRARk.

Studies have been done about the optimization of RAR password recovery algorithm on GPU. Andrew Adinetz et al. [2] find efficient methods for finding collisions for SHA-1 hashing on GPU by reducing branch divergence. David Apostal et al. [3] conduct the algorithm by multi-GPUs, and use MPI to minimize the amount of latency and handle the communication between the devices. They make the algorithm 57x and 40x faster than the serial program using 8,604,880 dictionary words. Pham Hong Phong et al. [4] reduce the password search space by using the password structure analysis technique for password recovery of ZIP. Qinjian Li et al. [5] store T-boxes in shared memory, leading to better throughput for AES encryption. Keisuke Iwai et al. [6] find that the decision of granularity and memory allocation is the most important factor for effective processing in AES encryption on GPU, but don't do much research on the decryption. Pedro Miguel Costa Saraiva [7] begins to study on the AES algorithm for GPU acceleration in 2013, and implements the AES algorithm in CBC mode and ECB mode. With the buffer size of 3.7MB, CBC decryption on a GPU can be up to 43 times faster than on the CPU.

Compared with the previous work, our work is different. Firstly, our work is aimed at recovering the password of an encrypted RAR file, whose algorithm is the combination of SHA-1 hashing and AES decryption. The two algorithms restrict with each other. Secondly, more complete and thorough optimizations are conducted for SHA-1 hashing in this paper, including the register usage optimization, and, therefore, vast performance improvement is achieved.

III. ALGORITHM INTRODUCTION AND PARALLEL ANALYZE

A. Algorithm Introduction

Because encryption of RAR files uses powerful AES-128 standard, which cannot be cracked. Therefore, dictionary file, brute force with mask and brute force are the only ways to recover the password of an encrypted RAR file. Fig.1 is the flow diagram of RAR password recovery algorithm. The following is a detailed explanation of the steps.

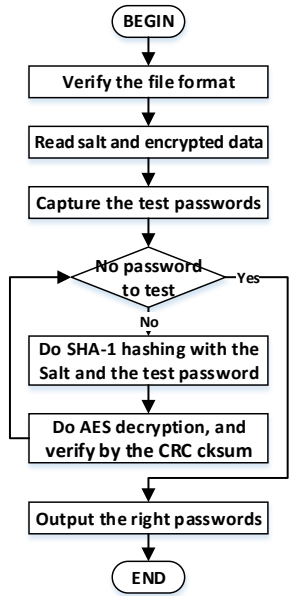

Fig. 1 The flow diagram of RAR password recovery algorithm

*1) Reading data from the encrypted RAR file to verify the file format:* The first 20 bytes of the encrypted RAR files are mark block and compressed file header, which can be used to verify the file format.

*2) Reading the Salt and encrypted file header block:* Salt is the encryption key, which will be used in SHA-1 hashing. Since the length of file header is uncertain, we use enough space, 1024 bytes, to store it.

*3) Capture the test passwords:* In dictionary file, there is a file storing the test passwords. In brute force, we have the charset and the length of password. The test passwords are all the possible combinations of characters in the charset with the required length. In brute force with mask, we can use the mask to exclude some combinations.

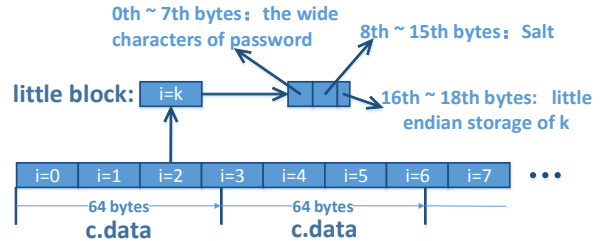

Fig.2 The assignment operation

*4) Do SHA-1 hashing with the Salt for every test password to get the array of key and initVector:* SHA-1 hashing is the most compute-intensive part in the process of RAR password recovery. To get the key and the initVector array, $4096 \times (password\_length \times 2 + 11)$ times of sha1_block() are needed, and before each execution of the sha1_block(), there are 64 times of assignment, and 192 times of condition statement. The process of SHA-1 hashing can be described as the pseudo-code blow:

```
Algorithm opt1: SHA1 hashing
1: Δ c is a SHA-1 data structure
2: Δ key and initVector are the input data of AES
3: unit_len ← password_len * 2 + 8 + 3
4: index ← 0
5: SHA1_Init(c)
6: for i←0 to 0x10 do
7:   SHA1_Final(c)
8:   Assign_init(initVector[i]; c.h4)
9:   for j ← 0 to 256*64*unit_len do
10:      Assign_data(c.data[index++]; k)
11:      if index = 64 then
12:         index ← 0
13:         sha1_block(c; c.data)
14:      endif
15:   endfor
16: endfor
17: SHA1_Final(c)
18: for i←0 to 16 do
19:   Assign_key(key[i];c)
20: endfor
```

When the length of password is 4, the work of assignment before the sha1_block() can be described in Fig.2. We use the little block to represent the data segment, which is composed of passwords, Salt and three bytes of the little block's index. Little block's length is $password\_length \times 2+8+3$ (unit_len will be used to represent this number blow). The "c.data" array is recurrently assigned by the little block.

*5) Do AES decryption for the encrypted file header, and verify the correctness by the CRC cksum:* When the length of

password is short, there can be collisions, so we still continue our test after getting a "right" one. The following pseudo-code represents this process: (In the following text, the calculation blow is immediately called AES decryption for short.)

```
Algorithm opt2: AES decryption and CRC cksum
1: Δ encryptBlock is the array of encrypted blocks
2: Δ decryptblock is the array of decrypted blocks
3: Δ each block's length is 16 bytes
4: Δ key and initVector are result of SHA-1 hashing
5: decryptblock[0]←AES1(encryptBlock[0],key)
6: decryptblock[0]←decryptblock[0]^initVector
7: for i←1 to number_of_blocks - 1
8:    decryptblock[i]←AES1(encryptBlock[i],key)
9:    decryptblock[i]←
          decryptblock[i]^encryptblock[i-1]
10: endfor
11: if getCRC(decryptblock; decryptblock) = true
12: then
13:    output the right password
14: endif
```

### B. Parallel Analyze and Naive Version

The calculations of all potential passwords are independent with each other, so the algorithm has good parallelism, and is suitable to work on GPU. Meanwhile, because the major calculation and time-consuming part, SHA-1 hashing, is hard to be parallelized, so this paper adopts coarse granularity parallel. The naive process of RAR password recovery algorithm on GPU can be separated to these three steps:

- Program reads the Salt and encrypted file header on CPU, and then transports them with the password set or the charset and length of password to GPU;

- The kernel executes on GPU. Each thread processes a test password, and records the password on global memory, if it's tested to be right;

- Then program runs back to CPU, reads the right passwords from GPU, and outputs them.

Because the test passwords could be a lot, for example, when the character set is all lowercase characters, and password length is 4, the number of all potential passwords is 456976. Testing them within a single execution is unrealistic. Considering resource allocation, GPU usually has a limit for the maximum number of threads in a single execution. For example, it is limited to 81920 on AMD HD 7970 [8]. In addition, if doing too much work in a single execution, GPU will continuously run for a long time without feedback, so the user cannot immediately see the program's running state and judge whether the GPU is stuck or still running. Besides, if the GPU fails during the execution, the entire process should be re-run, and it's obviously unreasonable to make a single execution work too much. Therefore, we separate the potential passwords to parts, and use a loop to test these parts sequentially. At the end of each iteration, we output the current test password, running speed, and the potential correct passwords.

In our naive version, SHA-1 hashing and AES decryption both execute on GPU in a same kernel. This version is simple but inefficient, since it has some performance problems, like the idle CPU, too much condition statements, the abuse of registers and so on. All these problems are dealt with step-by-step blow.

## IV. OPTIMIZATIONS

The kernel code of the naive version is similar with the CPU's, which is obviously unreasonable. GPU platform is very different from the CPU, and great performance can only be achieved by making full use of the GPU platform's advantages. Since this algorithm is compute-intensive, carefully optimizing the computing is significant. The following sections will detail the methods we utilize for optimal performance.

### A. Asynchronous Parallel Between CPU and GPU

Two parts of the calculation are SHA-1 hashing and AES decryption. The later requires the former's results, the key and the initVector. In the naive version, the two parts execute in a same kernel to reduce data transmission. Wherein, the computation amount of AES decryption is much smaller, relative to the SHA-1 hashing. The latter's run time is almost 270 times of the former's in the CPU serial version. But these few calculations may impact the kernel's parallelism.

AES decryption is data-intensive and needs lots of registers, which is discouraged for GPU program. Abuse of registers will limit the number of wavefronts working concurrently on the same compute unit [9]. When an active wavefront is blocked and switched for the ready wavefronts, registers used in that wavefront will stay on the compute unit. In this way, the switching of wavefronts can be fast. But concurrent wavefronts have to share the registers equally, which reduces the utilization of them. So less usage of registers is needed to make wavefronts work concurrently. Abuse of registers can also lead to the registers overflow. When the registers used in a thread are over the size of register vectors on the compute unit, for example 256 bytes on AMD Southern Islands architecture [8], some data originally stored in the registers will be dumped in the global memory. However, which part of data is moved to global memory is not controllable. If it happens to be the frequently accessed data, such as the wide characters of the password, the program performance will be greatly affected.

By analysis of the kernel program in the naive version, it can be found that the use of registers per thread is obviously excessive, and a large number of data in register are moved to global memory, resulting in performance degradation. Registers used in AES decryption account for about 80% of the total number. Thus, when coming to the algorithm optimization, the first consideration is to split the calculations to be two parts and move the AES decryption to CPU. Due to the small amount of AES decryption calculation, it can be executed fast on CPU and needn't to worry about these issues.

Furthermore, we use the asynchronous parallel between CPU and GPU to hide the time of AES decryption by the time of SHA-1 hashing. Asynchronous parallel implementation of CPU and GPU relies on the following functions: "clEnqueueNDRangeKernel", "clFlush" and "clFinish". "clEnqueueNDRangeKernel" enqueues tasks to command queue to wait for execution. "clFlush" pushes the tasks to GPU and makes sure they start to execute. The "clFinish" function is used to make sure the tasks in command queue are all finished [10]. Namely, the "clFinish" function is called after the "clEnqueueNDRangeKernel" function immediately, which leads to CPU idle in the GPU runtime. The asynchronous parallel is conducted by calling the "clFlush" function and AES decryption function before calling the "clFinish". In this way, the CPU and GPU can work simultaneously and take charge of different parts of the calculation. Thus, the computational resources can be fully used. Fig.3 is the flow chart of the asynchronous parallel, and Fig.4 is the comparison between the process of the synchronous parallel and the asynchronous parallel between CPU and GPU.

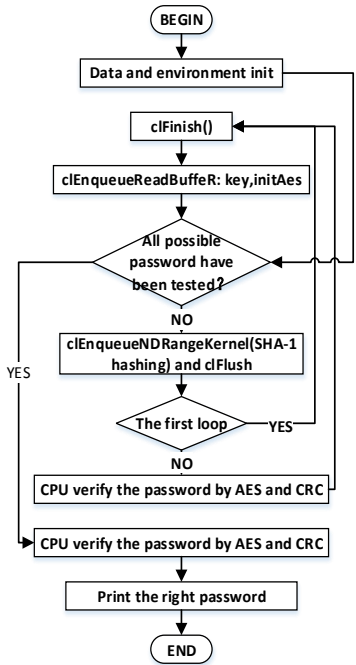 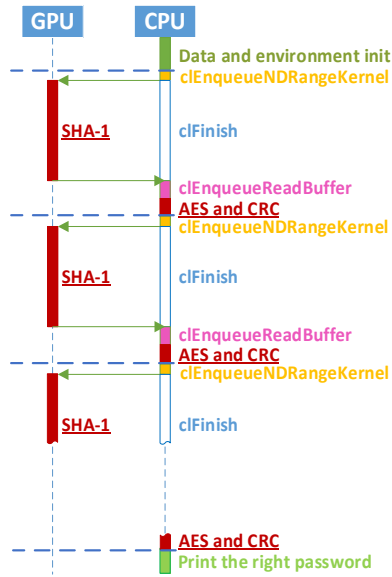 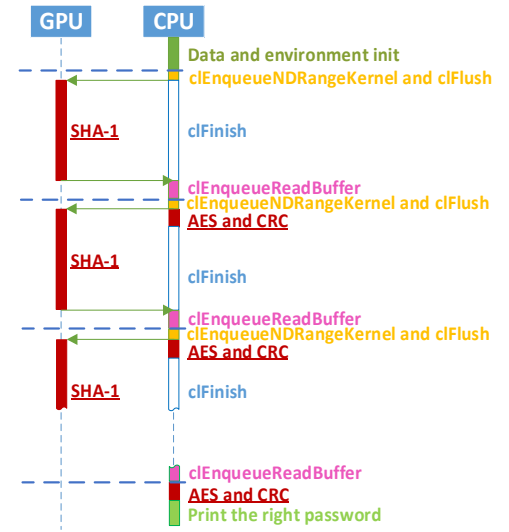

Fig.3 The flow chart of the asynchronous

Fig.4 Comparison between the synchronous and the asynchronous

Testing the run time of SHA-1 hashing on GPU and AES decryption on CPU with password length between 4 and 25 and passwords number per iteration 14336, it can be found that SHA-1 hashing is about 9~57 times slower than AES decryption. So the run time of AES decryption can be totally hidden by SHA-1 hashing. Detailed test data is shown in Fig.8.

### B. Redundant Calculations and Conditional Statements Reduction

As described in III.A, the two most time-consuming processes in the SHA-1 hashing are the sha1_block() function and the assignment before the sha1_block(). In the assignments, the 64 bytes array, "c.data", is repetitively assigned by the little block for 4096×unit_len times. The little block consists of the wide characters of the password, Salt and three bytes of the little block's index. When assigning the "c.data" array, wide characters of the password and Salt can be taken directly from the registers, while the three bytes of the little block's index has to be calculated, which is accumulated a large amount of computation. These calculations are exactly the same for the tests of all potential passwords, but now they need to execute repeatedly. SHA-1 hashing is already compute-intensive, and fetch almost has no occupancy of execution time. Therefore, we consider trading space for time. That is, calculating the data of little block's index on CPU, and storing them to "counter" array in the global memory, which is shared by all threads. In this way, redundantly calculating the data of little block's index is relieved. Threads can directly read the data from global memory.

On the other hand, because the size of "c.data" is not an integral multiple of the size of little block, and the little block consists of three different data sources, 3 conditional statements are needed for each byte's assignment, for the purposes of:

- *Update the index of bytes in the little block and the index of little block.* If the index of bytes in the little block is equal to unit_len, it should be set to "0" and the index of little block should plus one;

- *Update the index of bytes in the "c.data" array and determine the need for sha1_block() function.* If the index of bytes in the "c.data" array is equal to 64, it should be set to zero and the sha1_block() function should be conducted;

- *Select the data to be assigned.* According to the index of bytes in the little block, choose the data source to be assigned to the "c.data". The choice is based on the little block's data composition.

Because GPU is not good at logic control [11], it is not conducive to take advantage of GPU for high performance when so many conditional statements exist. In view of this problem, we allocate 4096×64×unit_len bytes for the "counter" array to unify the assignment and reduce the number of conditional statements. 4096×64×unit_len is the total of bytes to be assigned. In the "counter" array, only the part of the data of the little block's index is calculated and stored, and the rest store "0". Meanwhile, a unit_len bytes array in the register, the "data" array, is needed to store the wide characters of password and Salt. Similarly, the remaining three bytes of "data" array store "0". Then the assignment calculation becomes a simple and unified OR operation of these two arrays' element.

The source of data used to assign are confirmed, so no conditional statement is needed to select the data source. The number of conditional statements required for each byte's assignment is reduced to 2 times. Since the assignment operation is unified, it becomes more flexible, and easier to optimize to further reduce the number of conditional statements. Optimization measures taken in this paper are loop unrolling, and combining assignments.

*Loop unrolling:* For each assignment, it's necessary to update the index of bytes in the "c.data" array and determine the need for sha1_block() function. If the index of bytes in the "c.data" array reaches 64, it should be set back to zero, and sha1_block() function is needed to be conducted. In the paper, we avoid this judgment by unrolling the loop for 64 times. Thus, the code length increases, but the number of conditional statements per byte reduces to be 1.

*Combining assignments:* After loop unrolling, there are 64 assignment operations for each cycle. We combine every four assignments of byte to one assignment of unsigned integer, to speed up the assignment and reduce the number of conditional statements. But the length of "data" array is not a multiple of 4, so we expand it by 4 times to an unsigned integer array with unit_len elements, whose data is a combination of four former "data" arrays. In this way, number of conditional statements is reduced to a quarter time per byte.

After these steps, the number of the conditional statements during assignment is reduced to a quarter time per byte from the original 3 times per byte. The optimized assignment is shown in Fig.5. Taking into account the number of bytes to assign, the eliminated amount of conditional statements is enormous.

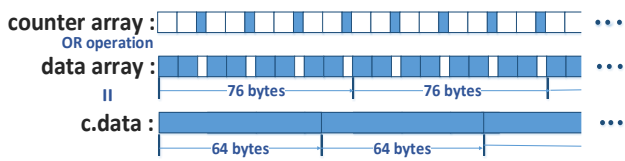

Fig.5 The optimized assignment

### C. the Usage of Registers Optimization

Since the calculation of SHA-1 hasing is massive, and difficult to be separated to work in parallel, so the RAR password recovery algorithm is coarse granularity parallel. In the calculation of SHA-1 hashing, a lot of registers are needed and the total amount is 356 bytes, when password length is 4. What is more, the amount will increase with the increase of password length. They mainly include the registers used in SHA-1 hashing and the "data" array.

As analyzed in Section A, there is a limit for the number of registers used in threads. If excessively using registers, some data supposed in registers will be dumped to global memory. For AMD FirePro W8000 GPU, each thread has up to 256 registers. The number of registers used in our program has exceeded the limit, and a lot of data is dumped to global memory. But what part of all these data is moved to global memory is uncontrollable and uncertain. To minimize the impact of excessive uasge of registers, in this paper, we store the frequently accessed data, the "data" array, to the local memory.

When dumping the "data" array to the local memory, each thread requires its own independent "data" array, and the length of the "data" array will increase with the increase of the password length. Take threads per workgroup 256 as an example, 256×unit_len×4 bytes of local memory are required to store the "data" array of all threads within a workgroup. But the local memory size is limited to 32KB, in AMD Radeon W8000, which means the maximum password length is 10, but this is clearly not enough. This paper tried two ways to handle this problem.

- Gradually reducing the number of threads within a workgroup when the password length is over 10. The optional and smallest number of threads within a workgroup is 64, so the maximum password length supported is 58. But we find that reducing threads per workgroup will clearly lead to performance degradation, so the number of threads per workgroup is the greatest possible number according to the password length in this paper.

- Reducing the data volume of "data" array to unit_len bytes when the password length is over 10. The supported maximum password length is the same with the former. In this way, we needn't reduce the threads per workgroup, which may cause more serious performance reduction.

After testing and comparing the two methods, the second one is much better than the first one. With the password length in the range of 11~25, the second method can achieve 1.1 ~ 1.2 times speed-up, compared with the first one.

### D. Others

Besides the optimizations above, we also try other approaches, which are described in details below.

*1) Reuse registers:* Since register resources occupied by variables cannot be released until kernel is finished, which is a waste of resources. This paper reuses registers to reduce the waste. Registers reusage is assigning two totally different meaning to one variable without affecting the correctness. It is detrimental to the readability of the program, but saves the registers and brings performance improvement.

*2) Optimization of calculations :* Division, multiplication and remainder are inefficient on GPU, so we shift them to the addition, subtraction and bit operations. For example, we use the bitwise-and operator to do the modulus operation. Besides, we use some build-in function, such as 'mul24()', to speed up the calculation. Since this algorithm is compute-intensive, the optimization of calculations can bring considerable performance improvements.

*3) Finding the most optimal number of workgroups:* This optimization is to determine the optimal number of workgroups by evaluating the program performance using different number of workgroups. The test data is shown in Fig.9. The more the number of workgroups is, the more likely the program perform fast. Because the times of kernel launch and synchronization will decrease with the increase of workgroups number. But meanwhile, each kernel's execution time will become longer, and intermediate results need to be waited for a longer time too. So the number should not be too large.

## V. PERFORMANCE EVALUATION

In this section, to evaluate our work, the performance of AccentRPR on CPU and the performance of our optimized GPU implementation are compared. Performance test data after each step of optimizations and the test data during the optimizations are also recorded, and analyzed in detail here. The way we measure performance is the number of passwords processed per second. We run the programs on Intel Core i5-3470 CPU and AMD Radeon W8000 GPU. Table 1 is the comparison of experimental hardware platform parameters.

TABLE.1 Comparison of Experimental Hardware Platform Parameters

|  | *AMD W8000* | *Intel i53470 CPU* |
|---|---|---|
| Processor main frequency | 0.88 GHz | 3.2GHz |
| The number of cores | 1792 | 4 |
| Peak Gflops | 3.23TFlops | 57.76GFlops |
| Memory Bandwidth | 176GB/s | 25GB/s |

### A. AccentRPR and Optimized GPU Implementation

AccentRPR is one of the most popular and efficient softwares to recover passwords of encrypted RAR files, and we use it as the benchmark to judge the quality of our work. Fig.6 is the performance of our final optimized version and its acceleration ratio, relative to the AccentRAR. As shown in Fig.6, our finial version achieves 43 ~ 57 times speed-up, compared with AccentRPR on CPU, which is a huge boost.

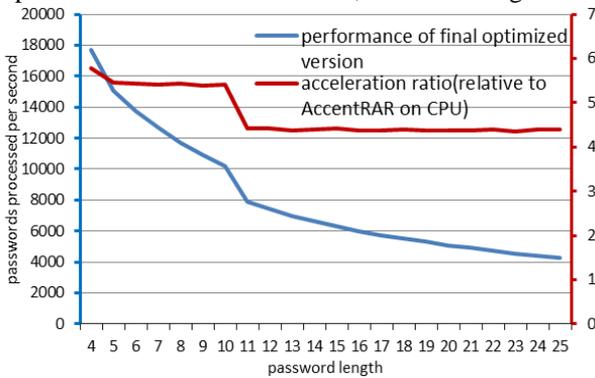

Fig.6 The performance of our final optimized version and the acceleration ratio

As shown in Fig.6, the performance decreases with the increase of password length, which is caused by the increase of the calculation amount. The number of loops of the most time-consuming calculation, sha1_block(), is 4096×unit_len. Unit_len is (password_length×2+11). So the calculation amount increases with the increase of password length. Besides, it can be found that the performance decreases greatly when the password length becomes 11. As we discussed in IV.D, because of the limit of local memory, we change the storage of the "data" array when the password length is over 11, which causes the increase of conditional statements and the decrease of performance.

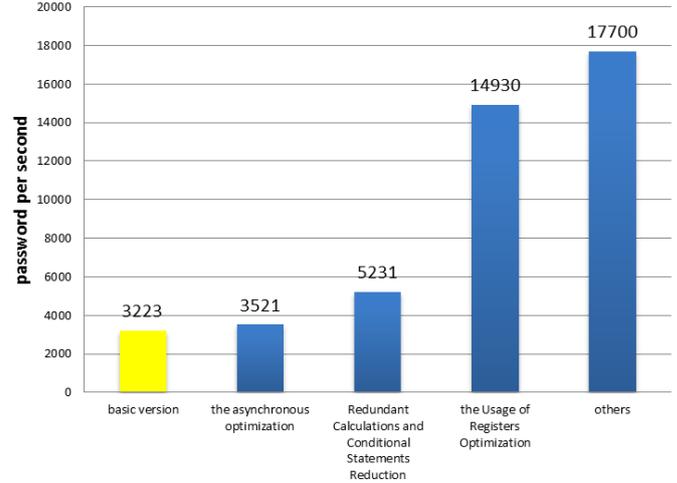

Fig.7 Performance after each step of optimizations

### B. Performance After Each Step of Optimizations

To assess the effect of optimizations, we record the performance after each step of optimizations with password length of 4, which is shown in Fig.7. Because AccentRPR supports GPU computing on AMD and NVIDIA graphics cards [13], we test the AccentRPR's performance on W8000 as the benchmark. It's obvious that redundant calculations and conditional statements reduction and the usage of registers optimization work very well.

After the asynchronous optimization, the performance is improved a bit, but lower than the expected. Because we believe that moving the verify calculation from GPU to CPU can save the usage of registers and improve parallelism of the kernel function, which has a great influence on performance. However, if we want to have wavefronts to concurrently execute on the same compute Unit to improve the parallelism, we should at least reduce the number of needed registers to 128, which is the half of the vector registers on a compute unit. Since the registers used in SHA-1 is 280 bytes, the parallelism doesn't improve. The improvement in this step is caused by the saving of the AES decryption run time by the asynchronous parallel optimization. To make sure the optimum effects of the asynchronous parallel, we test the run time gap between SHA-1 hashing on GPU and AES decryption on CPU, which is shown in Fig.8. According to the test result, run time of AES decryption can be completely hidden by SHA-1 hashing. Comparing the performance of before and after optimization, we can also speculate that the data dumped from register to global memory may mainly be the data in the AES decryption.

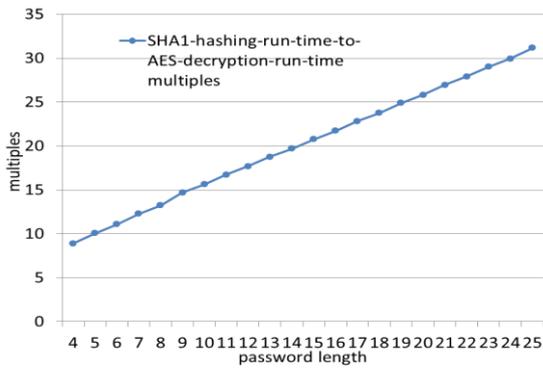
Fig.8 Run time comparison between SHA-1 hashing and AES decryption

The second step improves the performance for almost 1.5 times. This optimization measure brings more usage of registers and accessing to global memory, but it can greatly reduce the redundant calculation and the conditional statements. Since the GPU is not good at logic control, this change can make great acceleration.

Moving "data" array from register to local memory speeds up the kernel for about 2.8 times. In the previous version, the number of registers used in each thread is 356 when the password length is 4, plusing 10 for each addition of password length. Thus, some data are moved from register to global memory, because of the abuse of registers. But which part of data is moved is determined by the complier and it's user-uncontrollable. So it's advisable to move the frequently accessed data from register to local memory to avoid the case that these data are just moved to global memory. The "data" array is accessed frequently, and moving it from register brings such a great improvement, which can indicate that this array is moved to global memory in the previous version.

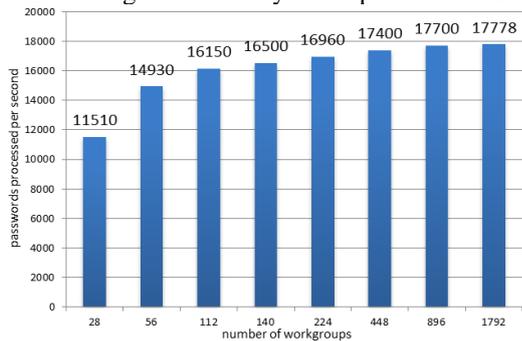
Fig.9 Performance test of number of workgroups

Some other optimizations like, reusing registers, optimization of calculations, finding the most optimal number of workgroups and so on are also conducted. Finding the most optimal number of workgroups seems to be the most effective in them. To find the most optimal number of workgroups, we try different number with the password length of 4, and the results are shown in Fig.9. Since W8000 has 28 compute units and great performance will be acquired when the number of workgroups is an integral multiple of the number of compute units [14], so we test 28, 56, 112, 140, 224, 448, 896, and 1792. According to the result, we can see that the performance improves with the increase of the number of workgroups, but grows slower and slower. The improvement is caused by the decrease of kernel launch and synchronization time. Considering the time of waiting for the intermediate results, we choose 896 to be the most optimal number of workgroups.

## VI. CONCLUSION AND FUTURE WORK

In this paper, we optimize the RAR password recovery algorithm by asynchronous parallel between CPU and GPU, redundant calculations and conditional statements reduction, the usage of registers optimization and others, like reusing registers. All these optimization methods contribute to the improvement of performance. The final version achieves a performance speed-up between 43 and 57 times, compared with AccentRAR on CPU. While, there are still some shortcomings in our algorithm, for example, the bad portability and the lack of optimizations for the test passwords search. Potential future work will focus on these problems.

ACKNOWLEDGMENT

We acknowledge the assistance of Ben Zhou in development process.